\documentstyle[epsf]{cupconf}
\title[Gravitational Dynamics in an Expanding Universe]{Gravitational Dynamics in an Expanding Universe}
\def\lb{\left(}
\def\rb{\right)}

\author[T. Padmanabhan]{T.\ls P\ls A\ls D\ls M\ls A\ls N\ls A\ls B\ls H\ls A\ls N\ls}

\affiliation{Inter-University Centre for Astronomy and Astrophysics,
Post Bag 4, Ganeshkhind, 
Pune - 411 007, INDIA.\\
 email: paddy@iucaa.ernet.in.}

\begin{document}
\ifnfssone
\else
  \ifnfsstwo
  \else
    \ifoldfss
      \let\mathcal\cal
      \let\mathrm\rm
      \let\mathsf\sf
    \fi
  \fi
\fi

\maketitle

\begin{abstract}
 The dynamical evolution of collisionless particles in an expanding
background is described. After discussing qualitatively the key
features, the gravitational clustering of collisionless particles in 
an expanding universe is modelled using some simple physical ideas. I
show that it is indeed possible to understand the nonlinear clustering
in terms of three well defined regimes: (1) linear regime (2)
quasilinear regime which is dominated by scale-invariant radial infall
and (3) nonlinear regime dominated by nonradial motions and
mergers. Modelling each of these regimes separately I show how the
nonlinear two point correlation function can be related to the linear
correlation function in hierarchical models. This analysis leads to
results which are in good agreement with numerical simulations thereby
providing an explanation for numerical results. The ideas presented
here will also serve as a powerful analytical tool to investigate
nonlinear clustering in different models. Several implications of the
result are discussed. 
\end{abstract}



\section{Introduction}

Consider a collection $N$ point particles, interacting with each other
by the Newtonian  gravity, in an expanding background characterized by
a scale factor $a(t)$. What can we say about the time evolution of
such a system? 

This problem is of considerable interest for several reasons. To begin
with, the behaviour of large number of particles interacting via
Newtonian gravity  poses a formidable challenge to the usual methods
of statistical mechanics (T. Padmanabhan, 1990). An isolated system,
made of such particles, possesses no equilibrium state in the
conventional thermodynamic sense. But -- as we shall see -- the
expanding background works in the direction opposite to the
gravitational clustering and changes the behaviour of the system
quantitatively. So, purely from an academic point of view, this seems
to be a challenging but solvable problem.

Secondly,  this problem might even have some practical interest. There
is considerable evidence that the universe is dominated by
collisionless non-baryonic dark matter particles. In that case, they
will play a key role in the formation of large scale structures. If
the length scales of interest are (i) small compared to  Hubble radius
but (ii) large compared to the  scales at which non-gravitational
processes are significant, then the system of dark matter particles
constitutes an example in which the question raised in the first
paragraph becomes relevant. In fact, most of the work in this subject
has been inspired by considerations of structure formation. 

A brute force method for solving this problem relies on numerical
simulations. In such an approach one starts with  large number of
particles distributed nearly uniformly and calculates the future
trajectories by a suitable numerical algorithm. Given a sufficiently
powerful computer, this procedure will lead to the positions and
velocities of the particles at any later epoch. All questions related
to this physical system can be answered using the output of the
numerical simulations and one may be tempted to declare the problem as
solved. 

I suspect  Donald Lynden-Bell will not quite like the above approach
to the solution of the problem. The ``Lynden-Bell tradition'' consists
of discerning the essence of the problem, modelling  it analytically
and obtaining  a solution which contains all the relevant
features. In this talk, I shall outline how one can make analytic
progress in the problem of gravitational clustering and thereby
reproduce the key features of numerical simulations. There are, of
course, some details which an analytic approach cannot bring out but
the analytic treatment has the advantage of providing genuine physical
understanding. 

In the next section I shall describe some general features of  the
gravitational clustering in an expanding background. Section 3 will
introduce the analytic model which is capable of reproducing the key
features of numerical simulations. The last section describes possible
extensions of this approach and conclusions. 

\section{Gravitational clustering in an expanding background}

Let us begin by considering the evolution of a system of particles
under  self-gravity in the absence of expansion. Such a system has no
stable thermodynamic equilibrium. It is possible to increase the phase
volume available for the system without bound by separating the
particles into a ``core'' and ``halo'' with the core becoming more and
more tightly bound and  the halo dispersing to larger and larger
radii. The ``final'' configuration for such a system will consist of a
few tightly bound binaries with the rest of the  particles dispersed
to large distances with positive energy. 

This situation changes drastically when we introduce an expanding
background. Consider a system of particles distributed homogeneously,
on the average, with a mean density $\bar \rho (t)$. This uniform
density will cause an expansion of the universe and the proper
distance ${\bf r}  = a (t) \bf x $ between particles will increase
with time. If the distribution was not strictly uniform, then the
perturbations in the density will act as local centres of
clustering. A region with overdensity will accrete matter around it
while an underdense region will repel matter in its surroundings. As a
result, perturbations in  density will tend to grow and when the
density contrast is of order unity, these cluster centres will exert
significant influence on the evolution. Particles in a highly
overdense regions will evolve essentially under  their own
self-gravity and will tend to form gravitationally bound
systems. Further evolution will crucially depend on a competition
between the influence of these clusters on each other compared to the
effects of expansion. The clusters themselves will attract each other
and can merge gravitationally to form still larger objects thereby
leading to a picture of hierarchical clustering. But the overall
effect of expansion will be to pull the cluster centres apart  thereby
reducing the effectiveness of mergers. 

In an underdense universe with $\Omega < 1 $ the exapansion will win
over merging at late times and we will be left with a bunch of stable,
virialized clusters flying apart from each other with the overall
expansion. Each of these individual systems will follow a core-halo
evolution leading to a tightly bound core and a dispersed halo. The
high energy halo will form a  hot background, punctuated by compact
gravitationally bound clusters. The evolution, in some sense,
``freezes out'' in such an $\Omega < 1 $ model. 

If $\Omega = 1 $, mergers of individual cluster centres can actively
compete with the effect of expansion. In such a case, it is far from
clear whether stable virialized clusters can exist over a Hubble
timescale. The evolution can go on forever hierarchically with mergers
and expansion delicately balanced.  

In the above description we have tacitly assumed that the initial
perturbation has power at all scales and - in particular - at small
scales. This allows clustering to proceed from small scales to large
scales. A somewhat different picture  emerges if the initial
perturbations have all their power concentrated on a narrow band of
width $\triangle L$ around some scale, say $L$. As the system evolves
we will first form objects with a characteristic scale $L$. At this
stage, the universe will be (essentially) made of shells with radius
$L$ and thickness $\triangle L$. But as time progresses, power from
large scales is  transferred to small scales. This small scale power
grows fairly fast and at late stages one has to take into account both
the initial power at large scales and the newly generated power at
small scales.

The above picture illustrates several interesting features of
gravitational clustering which any analytical model should take
cognizance of. To begin with, the transfer of power in gravitational
clustering is from large scales to small scales. This can be seen most
celarly from the following feature. Suppose, at some stage during the
evolution of clustering,  we lump together sets of nearby particles
and declare them as particles of larger mass.  We then continue the
evolution taking into account the gravitational interaction of these
new kind of particles. Such a procedure is equivalent to averaging
over smallscale power and it has very little effect on large
scales. [A more figurative way of saying this would be that highly
nonlinear density contrast inside a galaxy or a star has no effect on
the large scale dynamics of the universe]. But as the evolution
procees, one certainly generates small scale power from large scales
due to the breaking of long waves. In this sense gravitational
clustering is similar to fluid turbulence in which also we have
cascading of power from larger to smaller scales. 

The second point which is clear from the above discussion is that we
will most probably require two different kinds of approximations, one
to deal with hierarchical clustering and another one for models with
very little small scale power. In this talk  I shall concentrate on
models for hierarchical clustering and will make some  comments about
other  possibilities  in the last section. 

Thirdly, the qualitative picture of gravitational clustering shows
that one has to discover a suitable ``unit'' for the description of
the nonlinear phase. In the linear epoch one can study the evolution
very effectively using the Fourier components of the density
contrast. In other words, each of these Fourier modes is an
independent unit unaffected by the rest. In the highly nonlinear
epoch, the coupling between the modes cannot be ignored if one uses
standard Fourier transform techniques. It is, however, quite
conceivable that the nonlinear phase can be described by a
superposition of a different set of ``units''which evolve reasonably
independent of each other. For models with little  small scale power,
spherical shells with different radii and thickness seem to fit this
bill adequately. But a similar description for the case of
hierarchical clustering is hard to come by. We shall say more about
this in section 4.

\section{Model for nonlinear clustering}

When these density perturbations are small, it is possible to study
their evolution using linear theory. But once the density contrast
becomes comparable to unity, linear perturbation theory breaks down
and one must use  N-body simulations to study the growth of
perturbations. While these simulations are of  some value in making
concrete predictions for specific models, they do not provide clear
physical  insight into the process of non-linear gravitational
dynamics. To obtain such an insight into  this complex problem, it is
necessary to model the gravitational clustering of collisionless
particles using simple physical concepts. I shall develop one such
model in this section, which - in spite of extreme simplicity -
reproduces the simulation results for hierarchical models fairly
accurately. Further, this model also provides insight into the
clustering process and can be modified to take into account more
complicated situations. 

The paradigm for understanding the clustering is based on the well
known behaviour of a spherically symmetric overdense region in the
universe. In the behaviour of such a region, one can identify
three different regimes of interest: (1) In the early stages of the
evolution, when the density contrast is small, the evolution is
described by linear theory. (2) Each of the spherical shells with 
an initial radius $x_i$ can be parametersed by a mass contained inside
the shell, $M(x_i),$ and the energy, $E(x_i)$ for the particular
shell. Each shell will expand to a maximum radius $x_{max}\propto
M/|E|$ and then turn around and collapse. Such a spherical collapse
and resulting evolution allows a self similar description (Filmore \&
Goldreich, 1984; Bertshinger, 1985) in which each shell acts as though
it has an effective radius proportional to $x_{max}$. This will be the
quasilinear phase. (3) The spherical evolution will break down during
the later stages due to several reasons. First of all, non radial
motions will arise due to amplification of deviations from spherical
symmetry. Secondly, the existence of substructure will influence the
evolution in a non-spherically symmetric way. Finally, in the real
universe, there will be merging of such clusters [each of which could
have been centres of spherical overdense regions in the begining]
which will again destroy the spherical symmetry. This will be the
nonlinear phase. 

The description given above is sufficiently vague and sufficiently
well known that one may suspect it can not lead to any insight into
the problem. In particular, structures observed in the real universe
are hardly spherical. I will show that it is, however, possible to
model the above process in a manner which allows direct generalisation
to the real universe. 

To do this we will begin by studying the evolution of system 
starting from a gaussian initial fluctuations with an initial power
spectrum, $P_{in}(k)$. The fourier transform of the power spectrum 
defines the correlation function $\xi(a,x)$ where $a\propto t^{2/3}$
is the expansion 
factor in a universe with $\Omega=1$. It is more convenient for our
purpose to work with the
average correlation function inside a sphere of radius $x$, 
defined by
\begin{equation}
\bar{\xi}(a,x)\equiv {3\over x^3}\int^{x}_{0}\xi(a,y)y^2 dy  
\end{equation}
In the linear regime we have $\bar{\xi}_L(a,x)\propto
a^2\bar\xi_{in}(a_i,x)$. In the quasilinear and nonlinear regimes, 
we would like to have prescription which relates the exact $\bar
\xi$ to the mean correlation function calculated from the linear
theory. One might have naively imagined that $\bar\xi(a,x)$ should
be related to $\bar\xi_{L}(a,x)$. But one can convince oneself that
the relationship is likely to be nonlocal by the following analysis:

Recall that, the conservation of pairs of particles, gives an exact
equation satisfied by the correlation function (Peebles, 1980):
\begin{equation}
{\partial\xi \over\partial t}+{1\over ax^2}{\partial\over\partial
x}[x^2(1+\xi)v]=0\label{qpaircon}
\end{equation}
where $v(t,x)$ denotes the mean relative velocity of pairs at
separation $x$ and 
epoch $t$. Using the mean correlation function $\bar\xi$ and a
dimensionless pair velocity $h(a,x) \equiv - (v/\dot{a}x)$, equation
(\ref{qpaircon}) can be written as
\begin{equation}
({\partial\over\partial \ln a}-h{\partial\over\partial \ln x})\,\,\,
(1+\bar{\xi})=3h(1+\bar{\xi})
 \end{equation}
This equation can be simplified by first introducing the variables
\begin{equation}
A=\ln a,\qquad X=\ln x ,\qquad D(X,A) = \ln (1+\bar{\xi})
\end{equation}
in terms of which we have (Nityananda and Padmanabhan, 1994)
\begin{equation}
{\partial D\over\partial A}-h(A,X){\partial D\over\partial
X}= 3h(A,X)\label{qkey}
\end{equation}
Introducing further a variable
$F=D+3X$, (\ref{qkey}) can be written  in a remarkably simple form as
\begin{equation}
{\partial F\over \partial A}-h(A,X){\partial F\over\partial
X}=0\label{qfunrel}
 \end{equation}
The charecteristic curves to this equation - on which $F$ is a
constant - are determined by 
$(dX/dA)=-h(X,A)$ which can be integrated if $h$ is known. But note that
the charecteristics satisfy the condition 
\begin{equation}
F=3X+ D=\ln [x^3(1+\bar{\xi})]={\rm constant}
\end{equation}
or, equivalently,
\begin{equation}
x^3(1+\bar{\xi})=l^3\label{qxandl}
\end{equation}
where $l$ is another length scale. When the evolution is linear at all
the relevant scales, $\bar{\xi}\ll 1$ and $l\approx x$. As clustering
develops,
$\bar{\xi}$ increases and $x$ becomes considerable smaller than $l$.
It is clear that the behaviour of clustering at some scale $x$ is
determined by
the original {\it linear} power spectrum at the scale $l$ through the
``flow of information'' along the charesteristics.
This suggests that {\it we should actually 
try to express the true
correlation function $\bar\xi(a,x)$ in terms of the linear correlation
function $\bar\xi_L(a,l)$ evaluated at a different point}.

Let us see how we can do this starting from  the quasilinear regime.
Consider a region surrounding a density peak in the linear stage,
around which we expect the clustering to take place. It is well known
that density profile around this peak
can be described by
\begin{equation}
\rho(x)\approx\rho_{bg}[1+\xi(x)] 
\end{equation}
Hence the initial mean density contrast scales with the initial shell
radius $l$ as $\bar\delta_i 
(l)\propto\bar\xi_L(l)$ in the initial epoch, when linear theory
is valid. This shell will expand to a maximum radius of $x_{max}
\propto l/\bar\delta_i\propto l/\bar\xi_L(l)$. In  scale-invariant,
radial collapse, models 
each shell may be approximated as contributing with a effective radius
which is propotional to $x_{max}$. Taking
the final effective radius $x$ as proportional to $x_{max}$, the final
mean correlation function will 
be
\begin{equation}
\bar\xi_{QL}(x)\propto \rho\propto {M\over x^3}
\propto {l^3\over (l^3/\bar\xi_L(l))^3}\propto
\bar\xi_L(l)^3 
\end{equation}
That is, the final correlation function $\bar\xi_{QL}$ at $x$ is the cube of
initial correlation function at $l$ where $l^3\propto x^3
\bar\xi_L^3\propto x^3\bar\xi_{QL}(x).$ This is in the form demanded
by (\ref{qxandl}) if $\bar\xi\gg 1$. {\it Note that we did not assume that
the initial power
spectrum is a power law to get this result.} 

In case the initial power spectrum is a power law, with
$\bar\xi_{L}\propto x^{-(n+3)}$, then we immediately find that
\begin{equation}
\bar\xi_{QL}\propto x^{-3(n+3)/(n+4)}\label{qlndep}
\end{equation}
[If the  
correlation function in linear theory has the powerlaw form $\bar\xi_{L}
\propto x^{-\alpha}$ then the process desribed above changes the index
from $\alpha$ to $3\alpha/(1+\alpha)$. We shall comment more 
about this aspect later]. For the power law case, the
same result can be obtained by more explicit means. For
example, in power law models the energy of spherical shell will scale
with its radius as some power which we write as 
$E\propto x_i^{2-b}$. Since $M\propto x_i^3$, it follows that the
maximum radius reached by the shell scales as $x_{max}\propto
(M/E)\propto x_i^{1+b}$. Taking the effective radius as
$x=x_{eff}\propto x_i^{1+b}$,  the final density scales as
\begin{equation}
\rho\propto {M\over x^3}\propto {x_i^3\over x_i^{3(1+b)}}
\propto x_i^{-3b}\propto x^{-3b/(1+b)}\label{basres}
\end{equation}
In this quasilinear regime, $\bar\xi$ will scale like the density and we get
$\bar\xi_{QL}\propto x^{-3b/(1+b)}$. 
The index $b$ can be related to
$n$ by assuming the the evolution starts at a moment when linear
theory is valid. The gravitational potential energy [or the kinetic
energy] scales as $E\propto x_i^{-(n+1)}$ in the linear theory. This
may be seen as follows: The power spectrum for velocity field,
$P_v(k)$ in
the linear regime is related to that of density by $P_v\propto
P(k)/k^2\propto k^{n-2}$. Hence the contribution to $v^2$ in
each logarithmic scale in k-space is $k^3P_v/2\pi^2\propto k^{n+1}
\propto x^{-(n+1)}$. Similarly, the gravitational potential energy
due to {\it fluctuations} is
\begin{equation}
\phi\propto \int_0^x 4\pi y^2 dy{\xi(y)\over y}\propto x^2\xi(x)
\propto x^{-(n+1)}
\end{equation}
So the total energy in the initial configuration scales as
$x_i^{-(n+1)}$ allowing us to determine $b=n+3$. This 
shows that the correlation function in the quasilinear regime to
be the one given by (\ref{qlndep}) .
 
The case with power law initial spectrum has no intrisic scale, if $\Omega=1 $.
It follows that the evolution has to be self similar and
$\bar\xi$ can only depend on $q=xa^{-2/(n+3)}$. This allows to
determine the $a$ dependence of $\bar\xi_{QL}$ by substituting $q$
for $x$ in (\ref{qlndep}). We find
\begin{equation}
\bar\xi_{QL}(a,x)\propto a^{6/(n+4)}x^{-3(n+3)/(n+4)}\label{qlax}
\end{equation}
 Direct algebra shows that
\begin{equation}
\bar\xi_{QL}(a,x)\propto [\bar\xi_{L}(a,l)]^3\label{qlscal}
\end{equation}
reconfirming the local dependence in $a$ and nonlocal dependence
in spatial coordinate.
This result has no trace of original assumptions [spherical evolution,
scale-invariant spectrum ....] left in it and hence once 
would strongly suspect that it will have far general validity.

Let us now proceed to the third and nonlinear regime. If we ignore the
effect of mergers, then it seems reasonable that virialised systems 
should maintain their densities and sizes in proper coordinates, i.e.
the clustering should be ``stable". This
would require the correlation function to have the form $\bar\xi_{NL}
(a,x)=a^3F(ax)$. [The factor $a^3$ arising from the decrease in
background density].
From our previous analysis we expect this to be a function of
$\bar\xi_L(a,l)$ where $l^3\approx x^3\bar\xi_{NL}(a,x)$. Let us write
this relation as
\begin{equation}
\bar\xi_{NL}(a,x)=a^3F(ax)=U[\bar\xi_L(a,l)]\label{qtr} 
\end{equation}
where $U[z]$ is an unknown function of its argument which needs
to be determined. Since linear correlation function evolves as
$a^2$ we know that we can write $\bar\xi_L(a,l)=a^2Q[l^3]$
where $Q$ is some known function of its argument. [We are using
$l^3$ rather than $l$ in defining this function just for future
convenience of notation]. In our case $l^3=x^3\bar\xi_{NL}(a,x)
=(ax)^3F(ax)=r^3F(r)$ where we have changed variables from 
$(a,x)$ to $(a,r)$ with $r=ax$. Equation (\ref{qtr}) now reads
\begin{equation}
a^3F(r)=U[\bar\xi_L(a,l)]=U[a^2Q[l^3]]=U[a^2Q[r^3F(r)]]
\end{equation}
Consider this relation as a function of $a$ at constant $r$. Clearly
we need to satisfy $U[c_1 a^2]=c_2a^3$ where $c_1$ 
and $c_2$ are constants. Hence we must have
\begin{equation}
U[z]\propto z^{3/2}.
\end{equation}
Thus in the extreme nonlinear end we should have 
\begin{equation}
\bar\xi_{NL}(a,x)\propto [\bar\xi_{L}(a,l)]^{3/2}\label{qnlscl} 
\end{equation}
[Another way deriving this result is to note that if $\bar\xi=
a^3F(ax)$, then $h=1$. Integrating (\ref{qkey}) with appropriate boundary
condition leads to (\ref{qnlscl}) .]
Once again we did not need to invoke the assumption that the
spectrum is a power law. If it is a power law, then we get,
\begin{equation}
\bar{\xi}_{NL}(a,x)\propto a^{(3-\gamma)}x^{-\gamma};\qquad 
\gamma={3(n+3)\over (n+5)} 
\end{equation}

This result is based on the assumption of ``stable clustering" and
was originally derived by Peebles (Peebles, 1965). It can be directly
verified that the right hand side of this equation can be expressed in
terms of $q$ alone, as we would have expected. 

Putting all our results together, we find that the nonlinear mean
correlation function can be expressed in terms of the linear mean 
correlation function by the relation:
\begin{equation}
\bar \xi (a,x)=\cases{\bar \xi_L (a,l)&(for\ $\bar \xi_L<1, \, \bar
\xi<1$)\cr 
{\bar \xi_L(a,l)}^3 &(for\ $1<\bar \xi_L<5.85, \, 1<\bar \xi<200$)\cr
14.14 {\bar \xi_L(a,l)}^{3/2} &(for\ $5.85<\bar\xi_L, \, 200<\bar
\xi$)\cr}\label{qtrial} 
\end{equation}
The numerical coefficients have been determined by continuity
arguments. We have assumed the linear result to be valid upto
$\bar\xi=1$ and the 
virialisation to occur at $\bar\xi\approx 200$
which is result arising from the spherical model.  The exact values of
the numerical coefficients can be  obtained only from simulations. 

\begin{figure}
\epsfxsize=4.8truein\epsfbox[1 360 430 716]{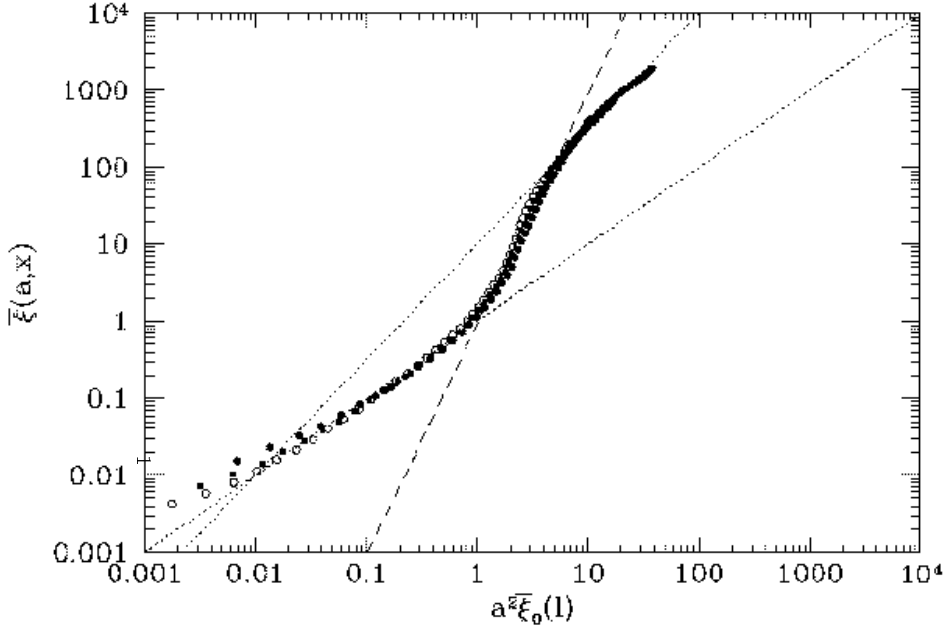}
\caption{Plot of $\bar\xi(a,x)$ against $\bar\xi_{L}(a,l)$ for CDM model.
The slopes in the three different regimes are indicated. The data points
are for three different redshifts [0.1,0.5 and 1.0] and are based on 
the simulations described in Padmanabhan et al ( 1995).}
\end{figure}

The true test of such a model, of course, is N-body simulations and
remarkably enough, simulations are very well represented by relations
of the above form.
Figure 1 shows the results of a CDM simulation based on the
investigations carried out in  Padmanabhan et al. (1995). This data
can be fitted 
by the relations (Bagla \& Padmanabhan, 1993):
\begin{equation}
\bar \xi(a,x)=\cases{\bar \xi_L(a,l) &(for\ $\bar \xi_L<1.2, \, \bar
\xi<1.2$)\cr 
{\bar \xi_L(a,l)}^3 &(for\ $1<\bar \xi_L<5, \, 1<\bar \xi<125$)\cr
11.7 {\bar \xi_L(a,l)}^{3/2} &(for\ $5<\bar\xi_L,  125<\bar
\xi$)\cr}\label{qbagh} 
\end{equation}
[The fact that numerical simulations show a correlation between
$\bar\xi(a,x)$ and $\bar\xi_L(a,l)$ was originally pointed out 
by Hamilton et al. (1991) who, however, tried to give a multiparameter
fit to the data. This fit has somewhat obscured 
the simple physical interpretation of the result though has the virtue
of being very accurate for numerical work.]

A comparison of (\ref{qtrial}) and (\ref{qbagh}) shows that the
physical processes 
which operate at different scales are well represented by our model.
In other words, the processes descibed in the quasilinear and nonlinear
regimes for an {\it individual} lump still models the {\it average}
behaviour of
the universe in a statistical sense. It must be emphasised that the key
point is the ``flow of information" from $l$ to $x$ which is an exact
result.  Only when the results of the specific model are recast in
terms of suitably chosen variables, we get a relation which is of general
validity. It would have been, for example, incorrect to use spherical
model to obtain relation between linear and nonlinear densities at
the same location or to model the function $h$. With hindsight, it is
clear why such attempts have not succeeded in the past.

It may be noted that to obtain the result in the nonlinear regime,
we needed to invoke the assumption of stable clustering which has
not been deduced from any fundamental considerations. In case
mergers of structures are important, one would consider this
assumption to be suspect (see Padmanabhan et al., 1995). We can,
however, generalise the above 
argument in the following manner: If the virialised systems have
reached  stationarity in the statistical sense, the function $h$
- which is the ratio between two velocities - should reach some
constant value. In that case, one can integrate (\ref{qfunrel}) and
obatin the result $\bar\xi_{NL}=a^{3h}F(a^hx)$. A similar argument
will now show that
\begin{equation}
\bar\xi_{NL}(a,x)\propto [\bar\xi_{L}(a,l)]^{3h/2}\label{qnlscl2}
\end{equation}
in the general case. For the power law spectra, one would get
\begin{equation}
\bar{\xi}(a,x)\propto a^{(3-\gamma)h}x^{-\gamma};\qquad 
\gamma={3h(n+3)\over 2+h(n+3)}
\end{equation}
Simulations are not accurate enough to fix the value of $h$; in
particular, the asymptotic value of $h$ could depend on $n$
within the accuracy of the simulations. It may be possible to
determine this dependence by modelling mergers in some simplified form.

If $h = 1$ asymptotically, the correlation function in the extreme
nonlinear end depends on the linear index $n$. One may feel that
physics at highly nonlinear end should be independent of the linear
spectral index $n$. This will be the case if the asymptotic value of
$h$ satisfies the scaling 
\begin{equation}
h = {3c \over n+3} 
\end{equation}
in the nonlinear end with some constant $c$. Only high resolution
numerical simulations can test this conjecture that $h(n + 3 ) = {\rm
constant}$.

It is possible to obtain similar relations between $\xi(a, x)$ and
$\xi_L (a, l) $ - in two dimensions as well. In this case the pair
conservation equation can be transformed to 
\begin{equation}
{\partial D \over \partial A} -h (A, x) {\partial D \over \partial x}
= 2h (A, X), 
\end{equation}
and the characteristics are determined by the relation
\begin{equation}
x^2 (1 + \bar \xi) = l^2 
\end{equation}
The self similar models due to Filmore and Goldreich (1984) show that
even in two dimensions $x_{\rm max} \propto l/\bar\delta_i \propto
l/\bar\xi_L (l)$. Repeating the previous analysis and noting that in
two dimensions $M \propto x^2$, we find $\bar\xi_{QL} (x) \propto
\left[ \bar\xi_L (l) \right]^2$. In the nonlinear limit if we invoke
stable clustering, then $\bar\xi_{NL} (a, x) =a^2 F(ax)$. An analysis
similar to the one performed before will now show that
$\bar\xi_{NL}(a, x) \propto \bar\xi_L(a, l)$. Thus in 2-D the scaling
relations are 

\begin{equation}
\bar \xi (a,x)\propto \cases{\bar \xi_L (a,l)&({\rm Linear}) \cr
\bar\xi_L(a,l)^2 &({\rm Quasi-linear})\cr
\bar\xi_L(a,l) &({Nonlinear}) \cr}
\end{equation}
For power law spectrum the nonlinear correction function will
$\bar\xi_{NL} (a, x) = a^{2 - \gamma} x^{-\gamma} $ with $\gamma = 2
(n + 2) / (n + 4)$. 

If we generalize the concept of stable clustering to mean constancy of
$h$ in the nonlinear epoch, then the correlation function will behave
as $\bar\xi_{NL} (a, x) = a^{2h}F(a^hx)$. In this case, if the
spectrum is a power law then the nonlinear and linear indices are
related to 
\begin{equation}
\gamma = {2h (n + 2) \over 2 + h (n + 2)}
\end{equation}
All the features discussed in the case of 3 dimensions are present
here as well. For example, if the asymptotic value of $h$ scales with
$n$ such that $h (n + 2 )= {\rm constant}$ then the nonlinear index
will be independent of the linear index. (Numerically it would be lot
easier to test this result in 2-D rather than in 3-D; work is in
progress to test these results). 

\section {Conclusions and Speculations}

The relations obtained  above will, of course, have certain limitations
on their validity. To begin with, we do expect a weak $n$-dependence
in these relations due to averaging over peaks of different heights. This
has been discussed using a simple analytic model, as well as numerically,
in  Padmanabhan et al. (1995). [Also see Mo et al., (1995) for a
similar discussion]. 
 Broadly speaking, we expect $h$ to be lower at a given $\bar \xi$ (in
the nonlinear regime with $\bar \xi > 10 $ or so ) as we add more
small scale power.          If stable clustering is invoked, then it
is possible to motivate this conclusion along the following lines: 
Consider a spherical
 region of initial radius 
$r_i$ and an 
overdensity of 
 $\nu \sigma$,
 where $\sigma = \sigma_0 r_i^{-(n + 3)/2}$ is the variance of the
gaussian density fluid. In a spherical model, this region will expand
to a maximum radius of about $( r_i / \nu \sigma )$ and virialise to a
final radius $r \equiv \lambda (r_i / \nu \sigma ) = ( \lambda / \nu
\sigma_0 ) r_i ^{n + 5 \over 2} $ where $\lambda \approx 0.5$. 
We shall assume that the coorelation function at the nonlinear end 
$\xi_{NL} (r)$
is contributed by such virialised objects and can be computed as
\begin{equation}
 1 + \xi (r) \approx \xi (r) = \langle \lb {r_i \over r } \rb^3 \rangle.
 \label{qrirav}
\end{equation}
[This assumption is equivalent to the stable
 clustering hypothesis.]
 From the relation $r = ( \lambda / \nu \sigma_0) r_i^{(n + 5)/2}$,
it follows that
\begin{equation}
\xi (r) = \lb {\sigma_0 \over \lambda} \rb ^{6/(n + 5 )} r ^{-{3 (n +
3 ) \over (n + 5 ) }} \left \langle \nu ^{6/(n + 5 )} \right
\rangle. \label{qsinl} 
 \end{equation}
Assuming that $\nu $ is a gaussian variable, we get

\begin{equation}
\langle \nu^{6/(n + 5 )} \rangle = {1 \over \sqrt{2 \pi} } 2 ^{{ 1 -n
\over 2n + 10 }} \Gamma \lb {n + 11 \over 2n + 10 } \rb. \label{qnuav}
\end{equation}
Equations (\ref{qsinl}  and  (\ref{qnuav}) show that $\xi (r) \propto
\xi^{3/2}_L [l]$ with $l \approx r \xi^{1/3}$; however, the
proportionality constant has a weak n-dependence. 

The averaging in (\ref{qrirav}) can be made more
sophisticated by using a weightage proportional 
to $r^m_i$. In that case, we still obtain the same $r$ dependence but
the proportionality constant becomes 
\begin{equation}
c = {\Gamma \lb {n + 11 + 2m \over 2n + 10} \rb \over \Gamma \lb
{n+5+2m \over 2n+10 }\rb}. 
\end{equation}
This reduces to the above result when $m = 0;$ recently, Mo et
al. (1995) have suggested a model based on $m = 3 $, which -- of
course -- leads to similar conclusions.

 Secondly, the asymptotic
behaviour will be sensitive to the value of $\Omega$. When $\Omega<1$,
structures ``freeze out" during the late stages of evolution and 
``stable clustering" is likely to be a reasonable assumption.

The description presented above is most suitable for models with
hierarchical clustering. As is well known, models like HDM which have
very little small scale power evolve quite differently. As an extreme
example of this case one may study the evolution of a model which has
power peaked at some scale $L$ with a band of width $\triangle
L$. During the linear epoch the power spectrum retains its shape and
evolves in a self similar manner. [This statement is not precisely
true. If the power spectrum falls faster than $\lambda^{-4}$ for large
$\lambda$, then the discreteness of particles will generate a
$\lambda^{-4}$ tail fairly rapidly. This effect, however, is important
only at large scales and we will be concerned with small scales.] At
some epoch, the scale $L$ will go nonlinear and we will form
structures which look typically like shells of radius $L$ and width
$\triangle L$. As evolution proceeds, small scale power is rapidly
generated due to the instability of the matter in the shell. At late
epochs, there is more power at small scales than at large scales due
to this effect. [ See figure 2.] 

\begin{figure}
\epsfxsize=4.8truein\epsfbox[15 402 555 750]{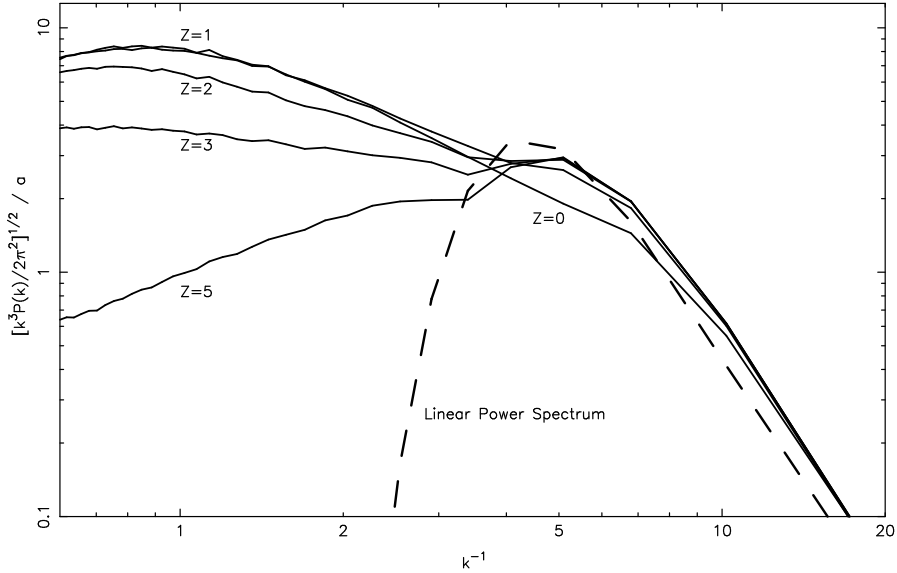}
\caption{Evolution of power spectrum for a model with power concentrated 
in a small range of scales. The dashed curve shows the linear power spectrum,
other curves show nonlinear power spectrum at five redshifts, scaled
to the final epoch. Sudden increase in power at small scales and later
saturation at a large value is apparent from this graph.}  
\end{figure}

Since the small scale power in the linear theory is exponentially
small, the nonlinear correlation function should be a rapidly growing
function of $\xi_L$ in order to reproduce the results of
simulation. This shows that the approach developed in the previous
sections is not very useful for HDM like models. It is however
possible to reproduce the simulation results of such models along the
following lines. 
 Consider an individual unit with a density profile $f
 ({\bf x }) = f (|{\bf x }|)$. If we now build a universe by
superposing several such clumps, then the density at any location will
be given by 
\begin{equation}
\rho({\bf x }) = \sum\limits_i f ({\bf x} - {\bf x}_i )
\end{equation}
It is easy to see that the power spectrum due to such a distribution
is given by 
\begin{equation}
P({\bf k}) = P_{BG} ({\bf k}) |f_{\bf k} |^2
\end{equation}
where $P_{BG}({\bf k})$ is the power spectrum corresponding to the
distribution of centres of the clumps. That is, $P_{BG}({\bf k})$ is
the fourier transform of the two-point correlation function of the
individual clumps.  At  high redshifts, virialized structures have not
formed and one may think of the original set of particles as
distributed according to some power spectrum. As time goes on,
individual clumps with some density profile originate leading to
significant amount of $|f_{\bf k}|^2$. We may now think of
$P_{BG}({\bf k})$ as the power spectrum corresponding to the
distribution of power centres (say, the minimum of gravitational
potentials) and $|f_{\bf k}|^2$ as a modulation of this power
spectrum. As time goes on, transfer of power takes place from scale to
scale changing the definitions of $P_{BG}$ and $f_{\bf k}$. At late
stages if one can identify some approximate nonlinear units then it
will be possible to build a specific model. For HDM-like spectra, one
models the universe as made of a bunch of spherical shells with
different radii and shell thickness. The individual evolution of each
such unit can be studied using spherical model and leads to results
which are in qualitative agreement with the simulations (Bagla and
Padmanabhan, 1995a). 

These results suggest the following paradigm for nonlinear clustering:
One is interested in expanding the density field as 
\begin{equation}
\rho(a,{\bf x})=\int\prod_{i=1}^N dc_i A(c_i,a)f(c_i,{\bf x}) 
\end{equation}
where $f(c_i,{\bf x})$ is a function which depend on a set of $N$ parameters
$c_i$ and $A(c_i,a)$ determines the weightage to different scales.
[In linear theory, $f=\exp i {\bf k.x}$ and the $c_i$'s are just ${\bf k}$] 
As evolution proceeds, one can ask how $A$ evolves and use this
information to quantify the notion of power transfer. Of course, the important
thing is to decide on a set of functions $f$ which are intutively simple,
well matched to the actual evolution and easy to work with. 
[In the case of fluid turbulence, one can use the concept of eddies to
model the dynamics to some extent; in a way, we are looking
for a concept analogous to that of eddies in the case of gravity]. It is fairly
straight forward to perform such analysis with functions $f$ which have support
in a small band of wavelengths [like the fashionable "wavelets"] and obtain
the evolution of $A$'s either from simulations or in some analytic
approximation. 
But it is doubtful whether such an exercise is of any use unless one
can show that the basis functions $f$ are better tuned to the actual
dynamics 
than the plane waves. This requires a careful analysis of the fully nonlinear
equations nonperturbatively using different basis functions
(Padmanabhan, 1995). 
The distinction between hierarchical clustering and HDM models will
then only be in the choice of specific nonlinear units chosen to
describe the evolution. Hopefully such an approach like this will
provide greater insight into the problem of nonlinear clustering. 

Finally note that the radial, scale invariant infall described in
the quasilinear regime has the effect of changing the linear
correlation function $\bar\xi_L=x^{-(n+3)}=x^{-b}$ to the quasilinear
correlation function $\bar\xi_{QL}=x^{-3b/(1+b)}$. It is amusing to
ask what will be the effect of iterating this process N-times. It is 
easy to see that the index after N iterations can be expressed in the form:
\begin{equation}
\gamma_N={A_N b\over 1+B_N b};\quad A_N=3^N;B_N={3^N-1\over 2}\label{qfxpt}
\end{equation}
The fixed point, of course, is $\gamma_{\infty}=2$ which is the only
nontrivial fixed point for such an evolution [with the other, trivial,
fixed point being zero]. If one could model the evolution as repeated
application of this process, one would expect a continuum of scaling
relations with the evolution being driven to a singular isothermal
sphere. The quasilinear evolution does not
change the $x^{-2}$ profile, a result which was noted earlier in Bagla
and Padmanabhan (1995). It is 
not clear whether the clustering can indeed be modelled using (\ref{qfxpt}) .

\section {Acknowledgements}

I thank J.S. Bagla, D. Lynden-Bell, R. Nityananda,
J.P. Ostriker and P.J.E. Peebles for several useful discussions. I thank
my coauthors  R. Cen, J.P. Ostriker 
and F. Summers for permission to adapt a figure from our joint work.

\end{document}